\documentclass[aip, floatfix, reprint]{revtex4-1}

\usepackage[utf8]{inputenc}
\usepackage[T1]{fontenc}
\usepackage{mathptmx}
\usepackage{svg}
\usepackage{amsmath}
\usepackage{amssymb}
\usepackage{amsfonts}
\usepackage{graphicx}
\usepackage{bm}
\usepackage{multirow}
\usepackage{xcolor}
\usepackage{siunitx}
\usepackage{booktabs}
\usepackage{xr-hyper}

\usepackage{hyperref}
\hypersetup{
    colorlinks=true,
    citecolor=black,
    linkcolor=black,
    urlcolor=cyan,
}

\newcommand{\ciso}{$^{13}\mathrm{C}$ }
\newcommand{\cisosingle}{$^{13}\mathrm{C}_{single}$ }
\newcommand{\cisocomb}{$^{13}\mathrm{C}_{combined}$ }
\newcommand{\hiso}{$^{1}\mathrm{H}$ }
\DeclareSIUnit\angstrom{\text {Å}}

\begin{document}

\title{Impact of noise on inverse design: The case of NMR spectra matching}

\author{Dominik Lemm}
\affiliation{University of Vienna, Faculty of Physics, Kolingasse 14-16, AT-1090 Vienna, Austria}
\affiliation{University of Vienna, Vienna Doctoral School in Physics, Boltzmanngasse 5, AT-1090 Vienna, Austria}
\author{Guido Falk von Rudorff}
\affiliation{University Kassel, Department of Chemistry, Heinrich-Plett-Str.40, 34132 Kassel, Germany}
\affiliation{Center for Interdisciplinary Nanostructure Science and Technology (CINSaT), Heinrich-Plett-Straße 40, 34132 Kassel}
\author{O. Anatole von Lilienfeld}
\email{anatole.vonlilienfeld@utoronto.ca}
\affiliation{Departments of Chemistry, Materials Science and Engineering, and Physics, University of Toronto, St. George Campus, Toronto, ON, Canada}
\affiliation{Vector Institute for Artificial Intelligence, Toronto, ON, M5S 1M1, Canada}
\affiliation{Machine Learning Group, Technische Universit\"at Berlin and Institute for the Foundations of Learning and Data, 10587 Berlin, Germany}

\date{\today}

\begin{abstract}
Despite its fundamental importance and widespread use for assessing reaction success in organic chemistry, deducing chemical structures from nuclear magnetic resonance (NMR) measurements has remained largely manual and time consuming. To keep up with the accelerated pace of automated synthesis in self driving laboratory settings, robust computational algorithms are needed to rapidly perform structure elucidations. 
We analyse the effectiveness of solving the NMR spectra matching task encountered in this inverse structure elucidation problem by systematically constraining the chemical search space, and correspondingly reducing the ambiguity of the matching task. 
Numerical evidence collected for the twenty most common stoichiometries in the QM9-NMR data base indicate systematic trends of more permissible machine learning prediction errors in constrained search spaces. 
Results suggest that compounds with multiple heteroatoms are harder to characterize than others.
Extending QM9 by $\sim$10 times more constitutional isomers with 3D structures generated by Surge, ETKDG and CREST, 
we used ML models of chemical shifts trained on the QM9-NMR data to test the spectra matching algorithms.
Combining both \ciso and \hiso shifts in the matching process suggests twice as permissible machine learning prediction errors than for matching based on \ciso shifts alone. 
Performance curves demonstrate that reducing ambiguity and search space can decrease machine learning training data needs by orders of magnitude.
\end{abstract}

\maketitle
\section{Introduction}

\begin{figure}
    \centering
    \includegraphics[width=\linewidth]{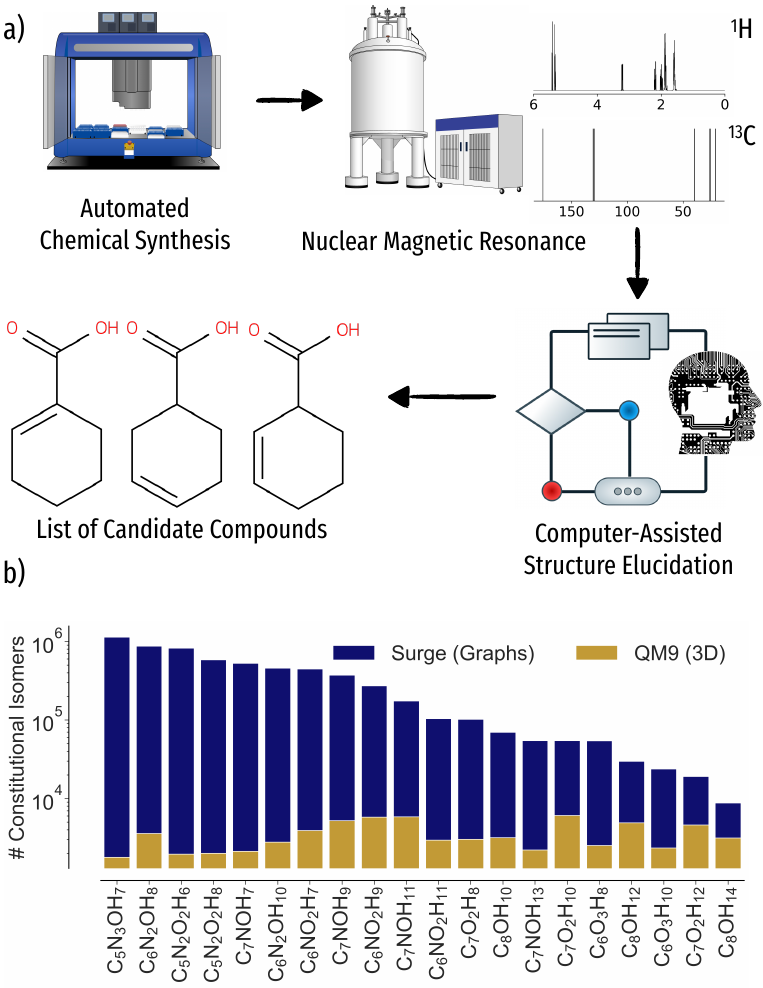}
    \caption{Schematic workflow for autonomous chemical discovery as well as scaling of constitutional isomer space versus data availability in the QM9\cite{ramakrishnan_quantum_2014} database.
    a) After the chemical synthesis of molecular compounds, reaction products are characterized using spectroscopic methods such as nuclear magnetic resonance (NMR). The measured \hiso and \ciso  spectra are automatically processed and potential candidate structures suggested via machine learning. 
    b) Number of constitutional isomers for 20 stoichiometries considered.
    }
    \label{fig:workflow}
\end{figure}

\begin{figure*}
    \centering
    \includegraphics[width=\textwidth]{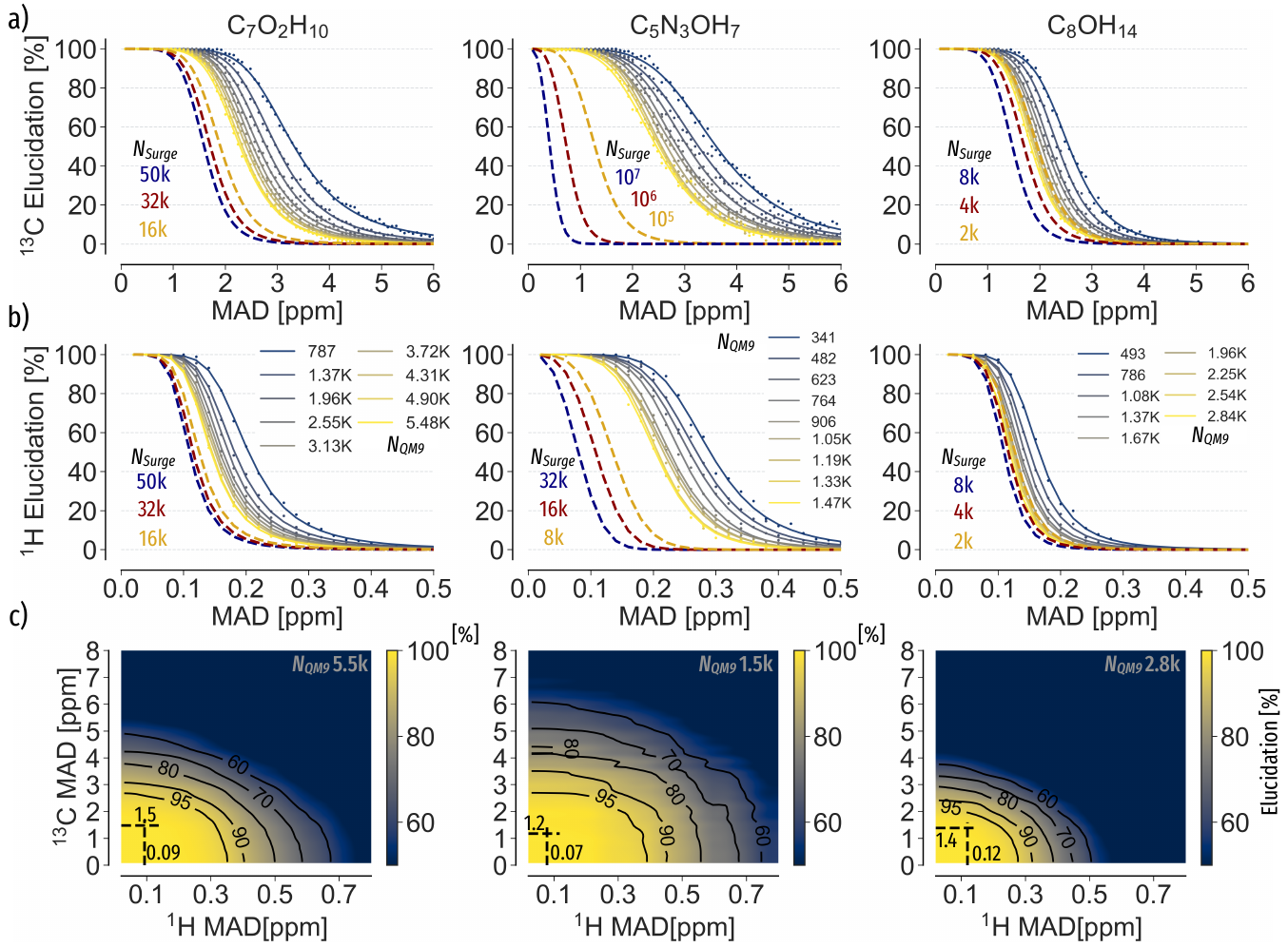}
    \caption{Elucidation performance curves of C$_7$O$_2$H$_{10}$, C$_5$N$_3$OH$_{7}$, C$_8$OH$_{14}$ spectra using Gaussian noise to control chemical shift accuracy in terms of mean absolute deviation (MAD) corresponding to $\sqrt{\frac{2}{\pi}}\approx0.8$ of the standard deviation\cite{Geary1935}.
    a-b) \ciso and \hiso spectra matching. Individual points were obtained by calculating the percentage of queries where noisy and noise free query spectra have the lowest distance. 
    All points have been fitted using Eq.\ref{eq:broken}. 
    Solid curves correspond to candidate numbers $N_{\rm QM9}$ from QM9\cite{ramakrishnan_quantum_2014}.
    Dashed curves are an extrapolation to candidate numbers $N_{Surge}$ as obtained via graph enumeration\cite{surge}.
    The legend corresponds to both a) and b), respectively.
    c) Spectra matching using both \hiso and \ciso shifts. 
    Dashed lines correspond to the accuracy required to correctly elucidate 95$\%$ of queries when only \hiso or \ciso spectra are being used, respectively.
    }
    \label{fig:synth_overview}
\end{figure*}

\begin{figure}
    \centering
    \includegraphics[width=\linewidth]{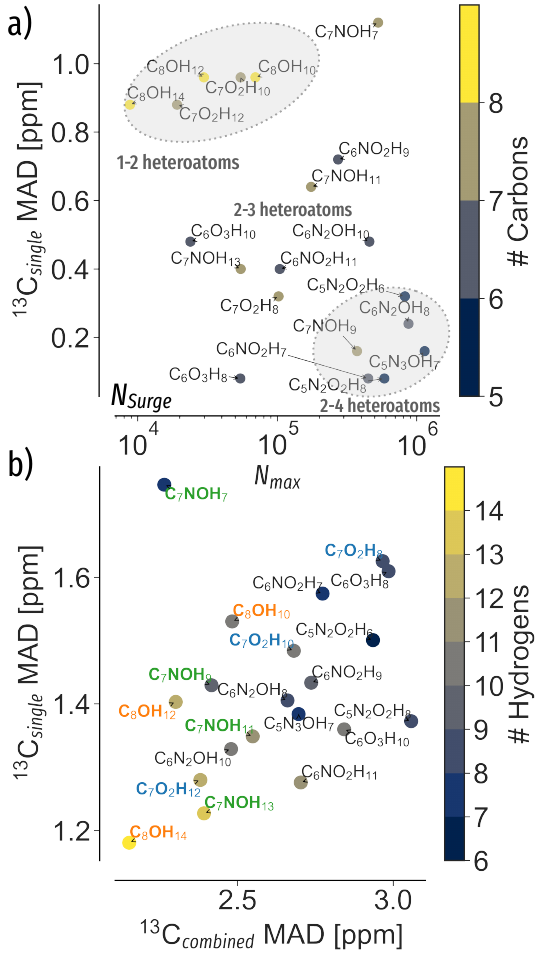}
    \caption{Trends in QM9\cite{ramakrishnan_quantum_2014} chemical compound space to correctly elucidate queries at 95$\%$ accuracy. The mean absolute deviation (MAD) $\sqrt{\frac{2}{\pi}}\approx0.8$ of the standard deviation\cite{Geary1935}.
    a) Extrapolated MAD at candidate numbers $N_{Surge}$ of the 20 most common stoichiometries in QM9\cite{ramakrishnan_quantum_2014}.
    b) MAD using only \ciso spectra (\cisosingle) against \ciso and noise-free \hiso spectra combined (\cisocomb) at candidate numbers $N_{QM9}$ from QM9\cite{ramakrishnan_quantum_2014}.
    }
    \label{fig:ccs}
\end{figure}

\begin{figure*}
    \centering
    \includegraphics[width=\textwidth]{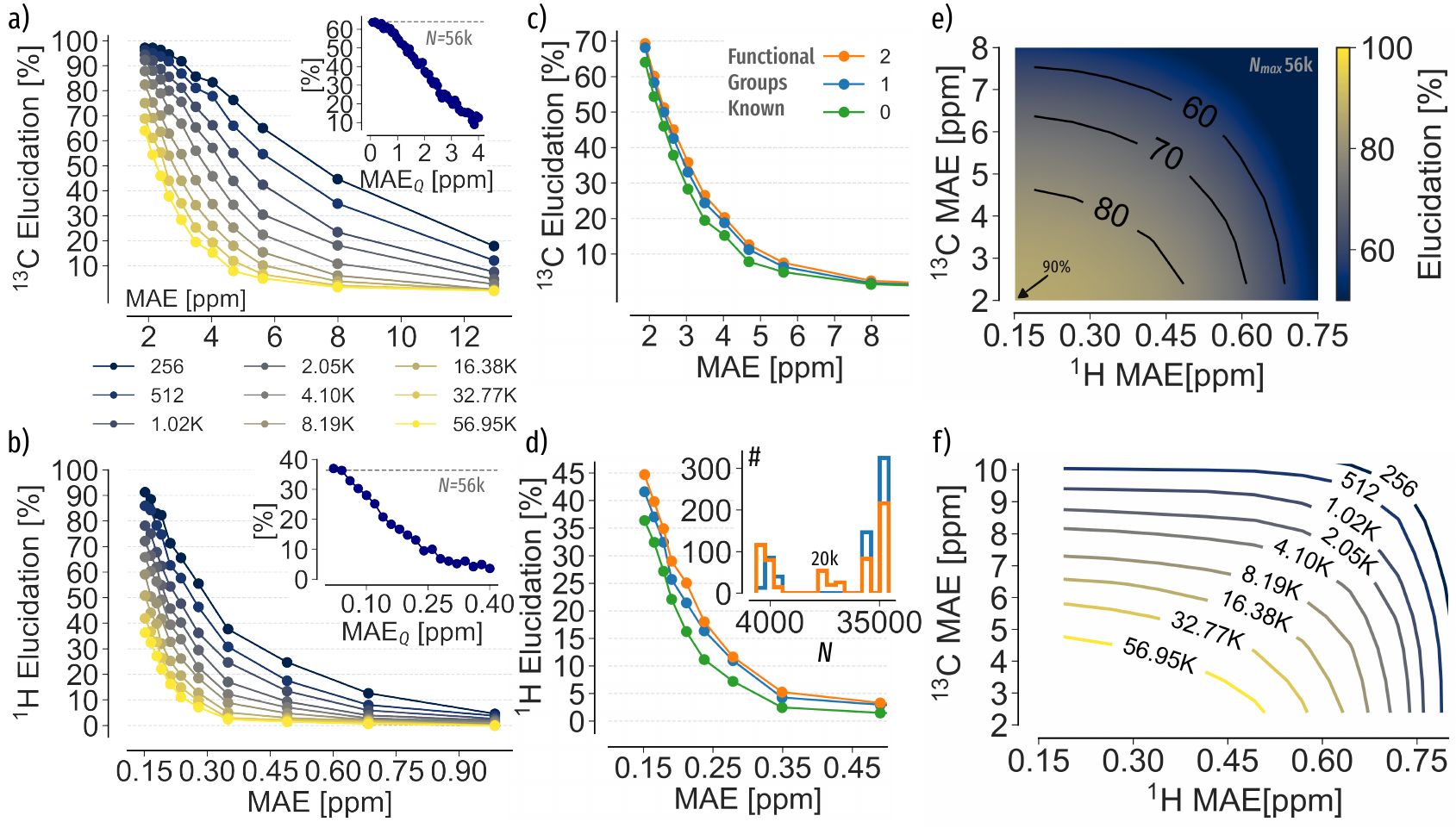}
    \caption{Elucidation accuracy of C$_7$O$_2$H$_{10}$ spectra using machine learning \ciso and \hiso shift predictions. Mean absolute error (MAE) refers to the predictive accuracy of the machine learning models, respectively.      
    a-b) \ciso and \hiso spectra matching at increasing search pool sizes $N$.
    The inset depicts the decay of the elucidation accuracy of the best performing machine learning model at increasing levels of Gaussian noise on query spectra (MAE$_Q$).
    c-d) Spectra matching accuracy when restricting the search pool to contain only known functional groups. The inset in d) depicts the search pool size $N$ restricted to compounds with similar functional groups as the query, respectively.
    e) Spectra matching using \hiso and \ciso shifts combined. 
    f) Accuracy required to reach 85$\%$ correct elucidation at increasing $N$ when using both \hiso and \ciso shifts combined.
    }
    \label{fig:ml}
\end{figure*}

Current development times of novel molecular materials can span several decades from discovery to commercialization.  
In order for humanity to react to global challenges, the digitization\cite{lab_automation22, chemdigitization19, GROMSKI20204, digisynth, EAST_review} of molecular and materials discovery aims to accelerate the process to a few years. 
Long experiment times severely limit the coverage of the vastness of chemical space, making the development of self driving laboratories for autonomous robotics experimentation crucial for high throughput synthesis of novel compounds (Fig.\ref{fig:workflow}~a))\cite{alanbayesian, XIE2023101043,jiang_airobots22, robotscientist, Burger2020, archemist22, closedloopopt_22}.
To keep the pace of automated synthesis, fast and reliable characterization of reaction products through spectroscopic methods is required, an often manual, time intense and possibly error prone task. 
One of the most common methods to elucidate the structure of reaction products are nuclear magnetic resonance (NMR) experiments.\cite{nmrrev15}
Through relaxation of nuclear spins after alignment in a magnetic field, an NMR spectrum, characteristic of local atomic environments of a compound, i.e. functional groups, can be recorded.
In particular, \hiso and \ciso NMR experiments are routinely used by experimental chemists to identify the chemical structure or relevant groups just from the spectrum.
For larger compounds, however, the inverse problem of mapping spectrum to structure becomes increasingly difficult, ultimately requiring NMR of additional nuclei, stronger magnets, or more advanced two-dimensional NMR experiments\cite{structureelucidator04, nmrchallenges17}.

Computer-assisted structure elucidation algorithms aim to iteratively automatize the structure identification process\cite{Willoughby2014-oh, case_review21, case20, dp4ai, dp5}.
Current workflows include repeated predictions of chemical shifts for candidate structure inputs through empirical or \textit{ab initio} methods\cite{hose, compNMR, Jonas2022-ff}.
Albeit accurate even in condensed phase through use of plane-waves~\cite{dsmp} or QM/MM setup~\cite{qmmmnmr}, the cost of density functional theory (DFT) calculations severely limits the number of candidate structures that can be tested, leaving the identification of unknown reaction products out of reach for all but the smallest search spaces.
Data driven machine learning models leveraging experimental or theoretical NMR databases\cite{nmrshiftdb, csd, qm9nmr, Bratholm2021} provide orders of magnitude of speedup over \textit{ab initio} calculations, reaching 1-2 ppm \ciso mean-absolute-error (MAE) w.r.t. experiment or theory, respectively\cite{Rupp2015-ds,nmrpred_kwon20, Jonas2019-oy, Han_nmr22, qm9nmr, Paruzzo2018, Musil2019}.
However, while the stoichiometry of the reaction product is usually known, e.g. through prior mass spectrometry experiments, the number of possible constitutional isomers exhibits NP hard scaling in number of atoms, quickly spanning millions of valid molecular graphs already for molecules of modest size (Fig.\ref{fig:workflow}~b)).
As such, the inverse problem of inferring the molecular structure from an NMR spectrum still poses a major challenge even for rapid solvers.

Recent machine learning approaches tackle the inverse problem using a combination of graph generation and subsequent chemical shift predictions for candidate ranking\cite{nmrinverse_jonas19, nmrinverse_huang21, nmrinverse_Sridharan22}.
First explored by Jonas\cite{nmrinverse_jonas19}, a Top-1 ranking with 57$\%$ reconstruction success-rate was achieved using deep imitation learning to predict bonds of molecular graphs.
Sridharan et al.\cite{nmrinverse_Sridharan22} used online Monte Carlo tree search to build molecular graphs resulting in a similar Top-1 ranking of 57.2$\%$.
Huang et al.\cite{nmrinverse_huang21} relied on substructure predictions from which complete graphs can be constructed, reaching 67.4$\%$ Top-1 accuracy by ranking substructure profiles instead of shifts.
A commonality between all algorithms is the subsequent ranking of candidates using spectra matching or other heuristics.
Consequently, even though the correct query compound could be detected early, similar candidates might be ranked higher, making the ranking process as critical as the candidate search itself. 

In this work, we analyse the effectiveness of the NMR spectra matching task encountered in the inverse structure elucidation problem.
As stagnating improvements\cite{Jonas2022-ff} in chemical shift predictions due to limited public NMR data aggravate candidate rankings, results suggest that both the prediction error of machine learning models \textit{and} the number of possible candidates are crucial factors for elucidation success.
By systematically controlling the size of chemical search space and accuracy of chemical shifts, we find that higher error levels become permissible in constrained search spaces.
Moreover, results indicate that increasing the uniqueness through including both \ciso and \hiso shifts in the matching process, rather than relying on a single type of shift, significantly reduces ambiguity and enhances error tolerance.
To evaluate the spectra matching task throughout chemical compound space, we systematically control the accuracy of 1D \ciso and \hiso chemical shifts of the 20 most common stoichiometries in QM9-NMR\cite{ramakrishnan_quantum_2014, qm9nmr} by applying distinct levels of Gaussian white noise.
Note that while we focus on DFT based 1D NMR in this work, future studies could include experimental data and 2D NMR information. 
Comparisons amongst stoichiometries suggest that chemical spaces with increasing amounts of heteroatoms and number of constitutional isomers are harder to characterize than others.
To test the spectra matching method on a large search space, we extended QM9-NMR to 56k C$_7$O$_2$H$_{10}$ constitutional isomers.
Controlling the chemical shift accuracy through machine learning models trained at increasing training set sizes, performance curves again indicate a trade-off between search space and accuracy.
Hence, as less accurate shift predictions become useful, results show that machine learning training data needs can be reduced by multiple orders of magnitude.

\section{Theory \& Methods}

\subsection{NMR Spectra Matching}

Consider a query \ciso or \hiso spectrum with a set of $N$ possible candidate constitutional isomer spectra.
We chose the squared euclidean distance as a metric to rank candidate spectra against the query spectrum (see SI Fig.3 for comparison against other metrics):

\begin{equation}
    d(\delta_{q}, \delta_{i}) = \sum_{j=1}^n (\delta_{q,j} - \delta_{i,j})^2,
\end{equation}

with $\delta$ being a sorted spectrum of $n$ chemical shifts (\ciso or \hiso), $q$ being the query, $i$ being the $i$-th of $N$ candidates, and $j$ being the $j$-th chemical shift in a spectrum, respectively.
To use both \ciso and \hiso shifts simultaneously for spectra matching, a total distance can be calculated as follows:

\begin{equation}
    d_{combined} = d(\delta^{13C}_{q}, \delta^{13C}_{i}) + \gamma \cdot d(\delta^{1H}_{q}, \delta^{1H}_{i}),
    \label{eq:comb}
\end{equation}

with $\gamma=64$ being a scaling factor determined via cross-validation (see SI Fig.1) to ensure similar weighting.
Final rankings are obtained by sorting all candidates by distance.
The Top-1 accuracy is calculated as the proportion of queries correctly ranked as the closest spectrum, respectively.

\subsection{Elucidation performance curves}

To analyse the spectra matching elucidation accuracy, we systematically control the number of possible candidates $N$ and the accuracy of chemical shifts, respectively.
For each constitutional isomer set, we choose 10$\%$ as queries and 90$\%$ as search pool, respectively.
Next, we randomly sample $N$ spectra from the search pool, including the query spectrum.
Each sample size is drawn ten times and the Top-1 accuracy averaged across all runs.
To control the accuracy of chemical shifts, we apply Gaussian white noise (up to 1 or 10 $\sigma$ for \hiso and \ciso, respectively) or use the machine learning error as a function of training set size (c.f. SI Fig.5 for learning curves).
For each $N$ and chemical shift accuracy, results are presented as elucidation performance curves, showing the elucidation success as a function of chemical shift accuracy in terms of mean absolute deviation (MAD) for Gaussian noise (c.f. Fig.\ref{fig:synth_overview}~a-b)) or mean absolute error (MAE) for machine learning predictions (c.f. Fig.\ref{fig:ml}).

\subsection{Chemical Shift Prediction}
We relied on kernel ridge regression (KRR) for machine learning \ciso and \hiso chemical shifts as presented in Ref.\cite{qm9nmr} and commonly being used in learning NMR properties from quantum chemical calculations\cite{Gerrard2020, Gerrard2021,gaumard2022regression,Tsai2022, Cordova2022, Paruzzo2018}.
We use a Laplacian kernel and the local atomic Faber-Christensen-Huang-Lilienfeld (FCHL19\cite{christensen_fchl_2020}) representation with a radial cutoff\cite{qm9nmr} of 4~$\si{\angstrom}$. 
The kernel width and regularization coefficient have been determined through 10-fold cross-validation on a subset of 10'000 chemical shifts of the training set. Note that while we relied on KRR within this work, other NMR shift estimation methods could have been used such as Hierarchically ordered spherical environment (HOSE) codes\cite{bremser1978hose, Kuhn2019} or neural network based approaches\cite{Unzueta2021, rull2023nmr, Han2021, jonas2019rapid}.

\subsection{Data}
The QM9-NMR\cite{qm9nmr, ramakrishnan_quantum_2014} dataset was used in this work, containing 130'831 small molecules up to nine heavy atoms (CONF) with chemical shieldings at the mPW1PW91/6-311+G(2d,p)-level of theory.
We used the 20 most common stoichiometries (Fig.\ref{fig:workflow}~b)), having a minimum of 1.7k constitutional isomers available in the dataset.

To extend the QM9-NMR C$_7$O$_2$H$_{10}$ constitutional isomers space, we 
used the systematic graph enumeration software Surge\cite{surge} to generate 54'641 SMILES. 
3D geometries of all SMILES have been generated using the ETKDG\cite{riniker_better_2015} method in RDKit. 
Lowest lying conformer structures were sampled using the CREST\cite{crest} algorithm, using the GFN2-xTB/GFN-FF composite method in a meta-dynamics based sampling scheme, with a final relaxation at the GFN2-xTB level.
Adding all successfully generated structures to QM9, a total pool size of 56.95k C$_7$O$_2$H$_{10}$ isomers was obtained.

For the training of chemical shift machine learning models, we selected  C$_8$OH$_{12}$, C$_8$OH$_{10}$, C$_8$OH$_{14}$, C$_7$O$_2$H$_{8}$ and C$_7$O$_2$H$_{12}$ constitutional isomers, yielding a total of 143k \ciso and 214k \hiso training points, respectively.

\section{Results \& Discussion}

\subsection{Spectra matching accuracy with synthetic noise}

To analyse the influence of noise and number of candidates on the elucidation success, we applied Gaussian noise to \ciso and \hiso shifts of C$_7$O$_2$H$_{10}$, C$_5$N$_3$OH$_{7}$ and C$_8$OH$_{14}$ constitutional isomers, respectively.
Fig.\ref{fig:synth_overview}~a-b) depicts a sigmoidal shaped trend of Top-1 elucidation performances as a function of mean absolute deviation (MAD) corresponding to $\sqrt{\frac{2}{\pi}}\approx0.8$ of the standard deviation\cite{Geary1935} caused by applying the Gaussian noise.
Note that increasing the maximum candidate pool size $N_{QM9}$ leads to an offset of the trend towards less permissible errors. 
A possible explanation is the correlation of the density of chemical space with increasing numbers of candidate spectra $N$\cite{SML}.
As shift predictions need to become more accurate, limiting $N$ through prior knowledge of the chemical space could be beneficial.
Similar findings have been reported by Sridharan et al.\cite{nmrinverse_Sridharan22}, noting that brute force enumerations of chemical space lead to worse rankings than constrained graph generation.
Note that while the trends in \ciso and \hiso elucidation are similar, less error is permissible when using \hiso shifts.

To further reduce the ambiguity, we include both \ciso and \hiso shifts into the matching problem as per Eq.\ref{eq:comb}.
Results suggest 50$\%$ and $\sim150\%$ more permissible \ciso and \hiso errors when both spectra are considered in the matching process (Fig.\ref{fig:synth_overview}~c)).
Similar to how chemists solve the elucidation problem, the inclusion of more distinct properties increases the uniqueness and can improve the elucidation success.

\subsection{Extrapolating the search space}

Due to the limited amount of constitutional isomers in databases compared to the number of possible graphs faced during inverse design (Fig.\ref{fig:workflow}~b)), assessing the chemical shift accuracy for successful elucidation is severely limited.
As such, we extrapolate elucidation performance curves to obtain estimates about chemical shift accuracies in candidate pool sizes larger than QM9.
We fit each elucidation performance curve (Fig.\ref{fig:synth_overview}~a-b)), respectively, using a smoothly broken power law function:

\begin{equation}
    f(x) = (1+ (\frac{x}{x_b})^d)^{\alpha}
\label{eq:broken}
\end{equation}

with $x_b$ controlling the upper bend and offset, $d$ changing the curvature and $\alpha$ changing the tilt of the function (see SI Fig.2), respectively.
The parameters of Eq.\ref{eq:broken} as a function of $N$ can again be fitted using a power law function (see SI Fig.2) and extrapolated to the total number of graphs $N_{Surge}$, respectively. 

Results of the extrapolation (Fig.\ref{fig:synth_overview}~a-b) dashed) indicate significant differences in elucidation efficiency among stoichiometries.
For instance, C$_8$OH$_{14}$ queries are potentially easier to elucidate than C$_5$N$_3$OH$_7$ structures.
Possible reasons are the limited number of C$_8$OH$_{14}$ graphs compared to millions of C$_5$N$_3$OH$_7$ isomers.
Moreover, the number of heteroatoms of the C$_5$N$_3$OH$_7$ stoichiometry might hamper the characterization when only relying on \ciso or \hiso, respectively.
Hence, to solve the inverse structure elucidation problem using experimental data of compounds larger than QM9, reducing ambiguities through including both \ciso and \hiso shifts as well as to reduce the candidate space is critical for elucidation success.

\subsection{Trends in chemical space}
To analyse the elucidation efficiency throughout chemical space, we applied the Gaussian noise and extrapolation procedure to the 20 most common stoichiometries in QM9 (Fig.\ref{fig:workflow}~b)).
Fig.\ref{fig:ccs}~a) shows the MAD required for 95$\%$ elucidation success as a function of $N_{Surge}$.
Results suggest that less error is permissible for stoichiometries with large $N_{Surge}$ and fewer carbon atoms.
As such, using only \ciso shifts might not be sufficient to fully characterize the compound.
Again, similar to how chemists use multiple NMR spectra to deduct chemical structures, additional information such as \hiso shifts are beneficial to extend the information content.

In Fig.~\ref{fig:ccs} b), the error permissiveness of spectra matching using only \ciso (see SI Fig.4 for \hiso) versus combining both \ciso and \hiso is being compared, revealing a linear trend between both.
Note that the C$_7$NOH$_7$ stoichiometry shows the smallest benefit from adding additional \hiso information.
Interestingly, a hierarchy for C$_7$NOH$_X$ stoichiometries of different degrees of unsaturation is visible, indicating an inverse correlation between number of hydrogens and \cisosingle MAD (Fig.~\ref{fig:ccs} b) green).
Similar hierarchies are also observed for other stoichiometries such as C$_7$O$_2$H$_{X}$ and C$_8$OH$_{X}$ (Fig.~\ref{fig:ccs} b) blue and orange).
On average, the combination of \ciso and \hiso for spectra matching increases the error permissiveness of \ciso and \hiso by 85$\%$ and 261$\%$ (see SI Fig.4), respectively.


\subsection{Comparison to machine learned shift predictions}

To test the elucidation performance using machine learning predictions, we trained \ciso and \hiso KRR models at increasing training set sizes (see SI Fig.5 for learning curves) and predicted chemical shifts of 56k C$_7$O$_2$H$_{10}$ constitutional isomers. Note that within this proof of concept application we rely on xTB-GFN2 relaxed geometries as queries, which on average are within 0.06~\si{\angstrom} RMSD of C$_7$O$_2$H$_{10}$ B3LYP level of theory structures\cite{Lemmg2s2021}.
Results again show similar trends as observed with Gaussian noise (Fig.\ref{fig:ml}~a-b)), however, indicate more permissive accuracy thresholds.
For instance, KRR \ciso predictions at 2~ppm MAE can identify 64$\%$ of queries rather than only 17$\%$ suggested by the Gaussian noise experiment.
The difference could be explained due the systematic, non uniform nature of the QM9\cite{ramakrishnan_quantum_2014} chemical space, influencing the shape and extrapolation of elucidation performance curves in Fig.\ref{fig:synth_overview}.
Moreover, Gaussian noise is applied to all shifts at random compared to possibly more systematic machine learning predictions.
Note that the trade-off between error and $N$ is consistent and that the exact parameters will depend on the machine learning model and the finite sampling of constitutional isomer space.

To model possible experimental noise on query spectra, we apply Gaussian noise to query spectra and evaluate the elucidation performance of the best performing machine learning model (see insets in Fig.\ref{fig:ml}~a-b)).
Results indicate a halving of elucidation accuracy when the query spectrum contains up to 2~ppm MAE$_Q$ in \ciso and 0.15~ppm MAE in \hiso error, respectively.
Thus, in the presence of experimental measurement noise even higher prediction accuracies might be necessary.
Combining both \ciso and \hiso spectra for matching improves the elucidation performance up to 90$\%$ (Fig.\ref{fig:ml}~e)).
Again, the combination of spectra for elucidation highlights the effectiveness of reducing the ambiguity of the matching problem by including additional properties.

Investigating potential strategies to reduce the constitutional isomer search space, we constrained $N$ based on functional groups (see SI Table~1).
Randomly selecting functional groups present in each query, $N$ can be reduced by 50$\%$ and 62$\%$ on average (see Fig.\ref{fig:ml}~d) inset for distributions), respectively.
Results in Fig.\ref{fig:ml}~c-d) indicate an increase of the elucidation accuracy by 5$\%$ in \ciso and up to 10$\%$ for \hiso, respectively, in agreement with the elucidation performance in Fig.\ref{fig:ml}~a-b).
Note that the knowledge of two functional groups only led to marginal improvements.
However, fragmentation could be more beneficial for larger compounds than present in QM9\cite{ramakrishnan_quantum_2014}, as reported by Yao et al.\cite{Yao2023}.
Using both \ciso and \hiso shifts on the reduced search space only lead to marginal improvements of 0.5$\%$ over the results of the full search space.

\begin{figure}
    \centering
    \includegraphics[width=\linewidth]{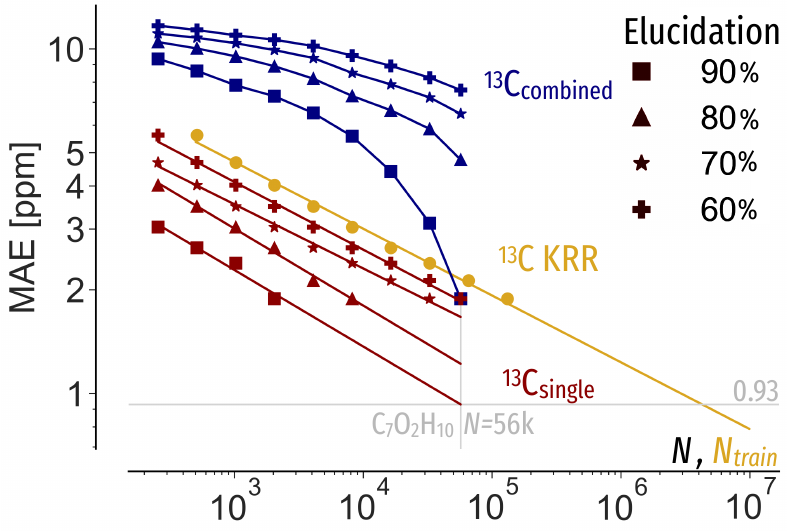}
    \caption{Performance curves (red, blue) of the MAE permissible to correctly identify 60, 70, 80, 90$\%$ of C$_7$O$_2$H$_{10}$ query spectra at a given pool size $N$ using machine learning shifts predictions, respectively.
    \cisosingle (red) only uses \ciso shifts for elucidation, whereas \cisocomb uses \ciso and \hiso spectra combined, assuming a \hiso MAE of 0.15~ppm.
    The learning curve (orange) indicates the systematic improvement of QM9\cite{ramakrishnan_quantum_2014} \ciso chemical shift predictions as a function of training set size $N_{train}$ using KRR with the FCHL19\cite{christensen_fchl_2020} representation.
    }
    \label{fig:pcurve}
\end{figure}

\subsection{Balancing search space and accuracy}

We use performance curves to analyse the relationship between the elucidation performance of C$_7$O$_2$H$_{10}$ queries, machine learning prediction errors and candidate pool sizes $N$.
Similar to learning curves, showing the systematic decay of out-of-sample machine learning prediction errors as a function of training data, elucidation performance curves show for a specific elucidation threshold, e.g. 90$\%$, the machine learning prediction error as a function of pool size.
Note that while learning curves of chemical shift predictions only show the predictive accuracy, e.g. in terms of MAE, the addition of elucidation performance allow a multifaceted evaluation of new spectra estimation algorithms, considering data efficiency as well as pool size.
The systematic decay of performance curves (Fig.\ref{fig:pcurve} red and blue) again demonstrates that constraining $N$ with prior knowledge allows for less accurate shift predictions to be applicable.
Extrapolating the \cisosingle performance curves indicates a machine learning MAE of 0.93~ppm to correctly rank 90$\%$ of queries out of 56k possible candidates (Fig.\ref{fig:pcurve} red), 0.02~ppm lower than suggested by Gaussian noise.
To reach an MAE of 0.93~ppm, four million training instances are required (Fig.\ref{fig:pcurve} orange).
Using both \ciso and \hiso shifts requires two orders
of magnitude less training data (Fig.\ref{fig:pcurve} blue).
As such, facing expensive experimental measurements and \textit{ab initio} calculations, more effective inverse structure elucidation could be achieved by balancing machine learning data needs through reduced search spaces and incorporation of additional properties.

\section{Conclusion}
We have presented an analysis of the effectiveness of the NMR spectra matching task encountered in the inverse structure elucidation problem.
By systematically controlling the predictive accuracy of \ciso and \hiso chemical shifts, we found consistent trends throughout chemical compound space, suggesting that higher errors become permissible as the number of possible candidates decreases.
Note that while we relied on 1D \textit{ab initio} NMR data, similar analysis could be performed using 1D or 2D experimental spectra.
Applications to the most common constitutional isomers in QM9 highlight that chemical spaces with many heteroatoms are harder to characterize when only relying on a single type of chemical shift.
Using both \ciso and \hiso chemical shifts increases the error permissiveness by 85$\%$ and 261$\%$ on average, respectively.
Machine learning predictions for 56k C$_7$O$_2$H$_{10}$ compounds showed that using both \ciso or \hiso shifts increased elucidation success to 90$\%$ compared to only 64$\%$ and 36$\%$ when used alone, respectively.
The usefulness of the analysis is expressed via performance curves, showing that training demands can be reduced by orders of magnitude compared to relying on specific shifts alone.

We believe that as the accuracy of machine learning models to distinguish spectra is limited, constrained search spaces or inclusion of more distinct properties are necessary to improve candidate rankings.
Rather than solely relying on more accurate models, future approaches could deal with estimating the applicability of machine learning models to successfully elucidate unseen chemical spaces, as well as including explicit knowledge of chemical reactions, functional groups or data from mass spectrometry, infrared- or Raman spectroscopy\cite{Gastegger2021, McGill2021, Grimme2013, Shrivastava2021, Jung2023, Pracht2020}, respectively.

Finally, explicitly accounting for atomic similarities and chemical shift uncertainties via the DP5 probability might further increase the confidence in structure assignments\cite{dp5}.

\section*{Acknowledgement}
O.A.v.L. has received funding from the European Research Council (ERC) 
under the European Union’s Horizon 2020 research and innovation programme (grant agreement No. 772834).
O.A.v.L. has received support as the Ed Clark Chair of Advanced Materials and as a Canada CIFAR AI Chair.
This research is part of the University of Toronto’s Acceleration Consortium, which receives funding from the Canada First Research Excellence Fund (CFREF).
Icons in Fig.\ref{fig:workflow} are from DBCLS, Openclipart and Simon Dürr from bioicons.com under CC-BY 4.0 and CC0, respectively.

\section*{Data \& Code Availability}
The QM9-NMR\cite{qm9nmr} dataset is openly available at \url{https://doi.org/10.17172/NOMAD/2021.10.16-1}.
The code and additional data used in this study are available at \url{https://doi.org/10.5281/zenodo.8126379}.

\section*{Conflict of Interest}
The authors have no conflict of interest.

\section*{References}
\bibliographystyle{vancouver}
\bibliography{references.bib}{}




\end{document}


\onecolumngrid

\begin{center}
  \textbf{\large Impact of noise on inverse design: The case of NMR spectra matching} \\[.5cm]
  \textbf{\large Supplementary Information} 
  \\[.5cm]
  Dominik Lemm$^{1,2}$, Guido Falk von Rudorff$^{3,4}$ and O. Anatole von Lilienfeld$^{5,6, 7}$\\[.1cm]
  {
  \itshape ${}^1$ University of Vienna, Faculty of Physics,  Kolingasse 14-16, AT-1090 Vienna, Austria\\
  \itshape ${}^2$ University of Vienna, Vienna Doctoral School in Physics, Boltzmanngasse 5, AT-1090 Vienna, Austria\\
  \itshape ${}^3$ University Kassel, Department of Chemistry, Heinrich-Plett-Str.40, 34132 Kassel, Germany\\
  \itshape ${}^4$ Center for Interdisciplinary Nanostructure Science and Technology (CINSaT), Heinrich-Plett-Straße 40, 34132 Kassel \\
  \itshape ${}^5$Departments of Chemistry, Materials Science and Engineering, and Physics, University of Toronto, St. George Campus, Toronto, ON, Canada\\
  \itshape ${}^6$Vector Institute for Artificial Intelligence, Toronto, ON, M5S 1M1, Canada \\
  \itshape ${}^7$Machine Learning Group, Technische Universit\"at Berlin and Institute for the Foundations of Learning and Data, 10587 Berlin, Germany\\
  }
  ${}^*$Electronic address: anatole.vonlilienfeld@utoronto.ca\\
(Dated: \today)\\[1cm]
\end{center}

\twocolumngrid

\begin{table*}[ht]
\centering
\caption{Functional groups contained in the C$_7$O$_2$H$_{10}$ constitutional isomer chemical space and corresponding SMARTS patterns.}
\begin{tabular}{lc}
\hline
\textbf{Functional Group} & \textbf{SMARTS Pattern} \\
\hline
alkene & [CX3]=[CX3] \\
alkyne & [CX2]\#[CX2] \\
arene & [cX3]1[cX3][cX3][cX3][cX3][cX3]1 \\
alcohol & [\#6][OX2H] \\
aldehyde & CX3H1[\#6,H] \\
ketone & [\#6]CX3[\#6] \\
carboxylic acid & CX3[OX2H] \\
acid anhydride & CX3[OX2]CX3 \\
ester & [\#6]CX3[OX2H0][\#6] \\
ether & OD2[\#6] \\
enol & [OX2H][\#6X3]=[\#6] \\
phenol & [OX2H][cX3]:[c] \\
\hline
\end{tabular}
\label{tab:fgs}
\end{table*}

\begin{figure*}
    \centering
    \includegraphics[width=0.75\textwidth]{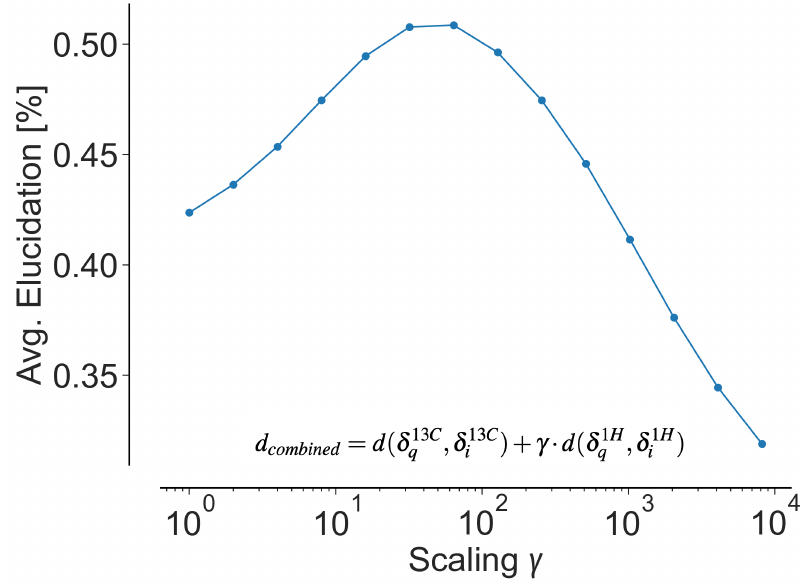}
    \caption{Hyperparameter scan of $\gamma$ on C$_7$O$_2$H$_{10}$ constitutional isomers for the combined ranking of \ciso and \hiso shifts. First, the respective distances of \ciso and \hiso at their individual shift accuracy levels are being calculated and then the distances combined via the depicted Eq.2.
    The average elucidation is calculated by averaging across all shift accuracy levels.}
    \label{fig:si_scaling}
\end{figure*}

\begin{figure*}
    \centering
    \includegraphics[width=\textwidth]{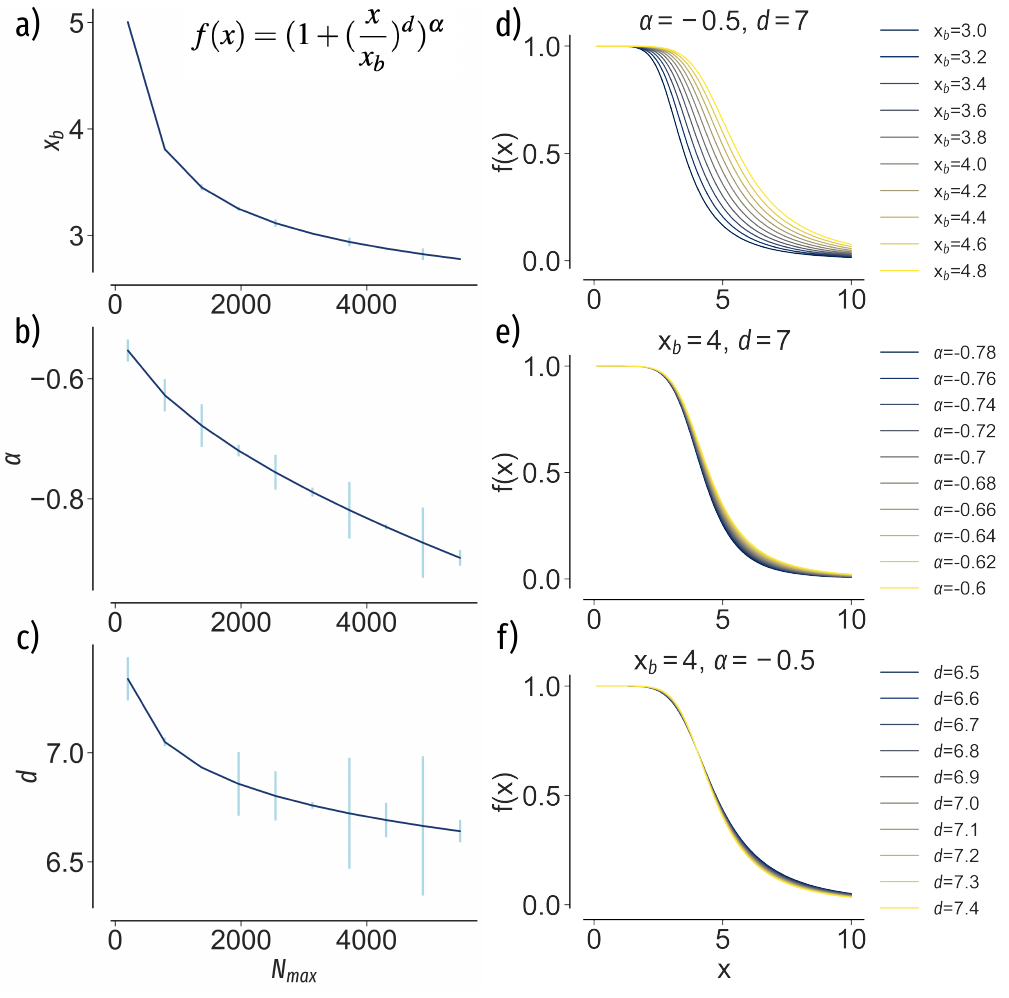}
    \caption{Parameter distributions of a broken powerlaw function (Eq.3) used for extrapolating the elucidation trends.
    a-c) Parameters $x_b$, $\alpha$ and $d$ fitted to the elucidation trends of C$_7$O$_2$H$_{10}$ at multiple $N_{max}$.
    Note that the parameters $d$ and $\alpha$ are more noisy in nature given the finite sampling and only marginally influence the shape of the curve in the observed parameter range (see e) and f)). 
    Conversely, the parameter $x_b$, which dictates the offset of the curve, is well behaved and decays smoothly as $N_{max}$ increases.
    d-f) Influence of the observed parameter ranges for $x_b$, $\alpha$ and $d$ on the shape of the broken powerlaw function.
    }
    \label{fig:si_fitting}
\end{figure*}

\begin{figure*}
    \centering
    \includegraphics[width=\textwidth]{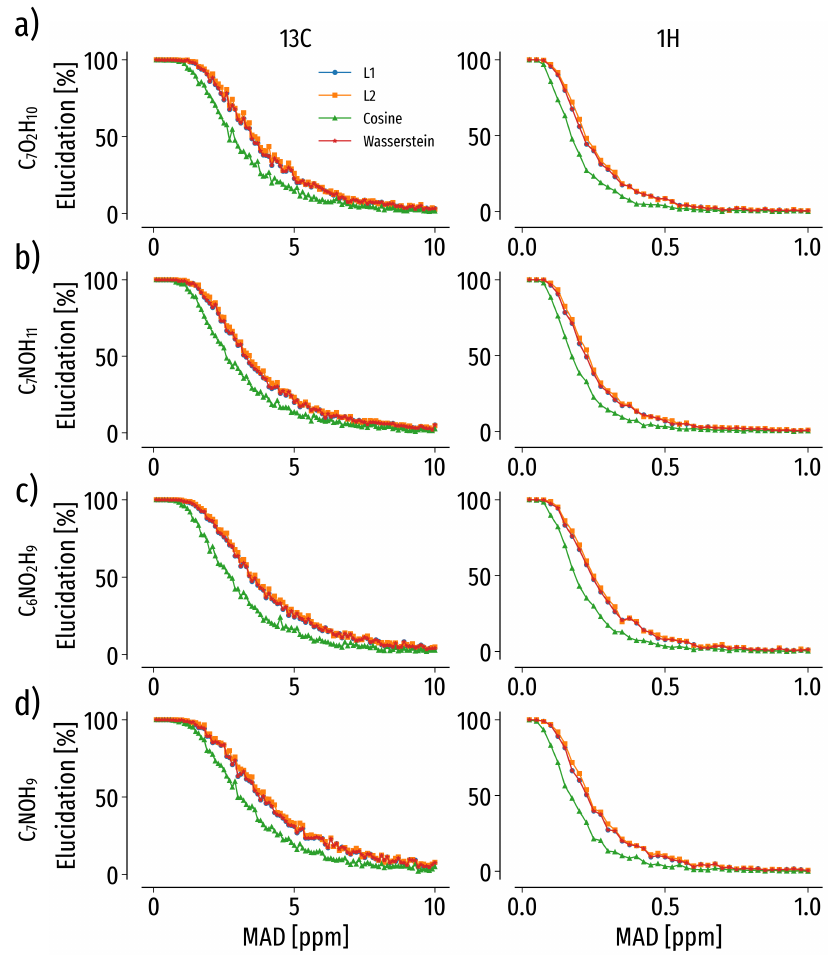}
    \caption{Comparison of L1, L2, cosine similarity and Wasserstein distances on the \ciso (left) or \hiso (right) elucidation success of C$_7$O$_2$H$_{10}$ (a), C$_7$NOH$_{11}$ (b), C$_6$NO$_2$H$_{9}$ (c) and C$_7$NOH$_{9}$ (d) constitutional isomers.}
    \label{fig:si_loss}
\end{figure*}

\begin{figure*}
    \centering
    \includegraphics[width=0.75\textwidth]{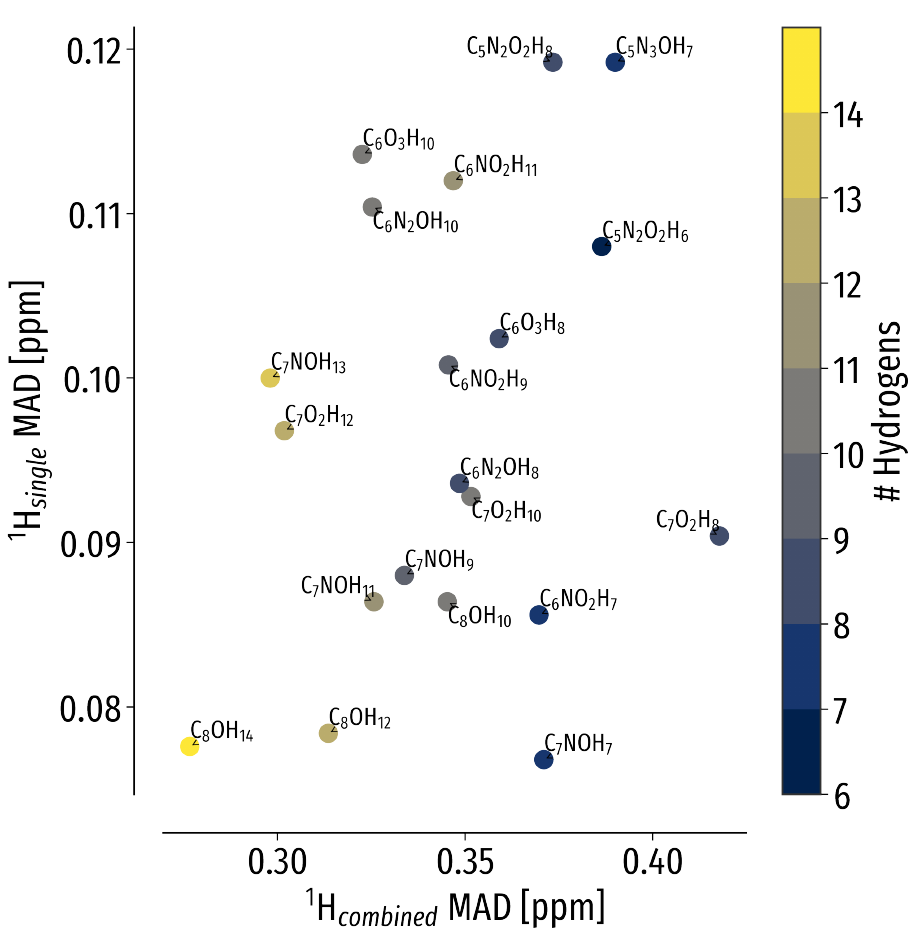}
    \caption{Trends in QM9 chemical compound space to correctly elucidate queries at 95$\%$ accuracy.
    Mean absolute deviation (MAD) using only \hiso spectra (\hiso$_{single}$) against \hiso and noise-free \ciso spectra combined (\hiso$_{combined}$) at the respective $N_{max}$ available in QM9.}
    \label{fig:si_hydro}
\end{figure*}

\begin{figure*}
    \centering
    \includegraphics[width=0.75\textwidth]{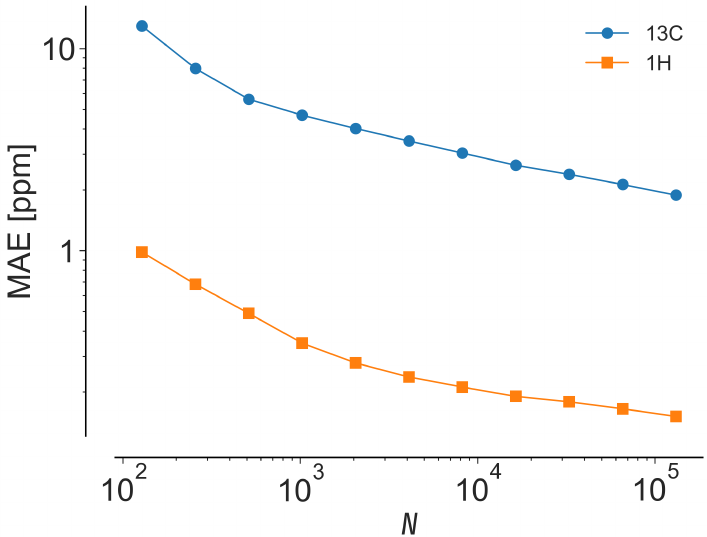}
    \caption{Systematic improvement with increasing training set size $N$ of KRR machine learning for \ciso and \hiso chemical shifts of C$_8$OH$_{12}$, C$_8$OH$_{10}$, C$_8$OH$_{14}$, C$_7$O$_2$H$_{8}$ and C$_7$O$_2$H$_{12}$ constitutional isomers using the FCHL19 representation.}
    \label{fig:si_learning}
\end{figure*}